\documentstyle[11pt]{article}
\newcommand{\be}{\begin{equation}}
\newcommand{\ee}{\end{equation}}
\newcommand{\bea}{\begin{eqnarray}}
\newcommand{\eea}{\end{eqnarray}}
\newcommand{\f}{\frac}
\newcommand{\e}{\epsilon}
\newcommand{\Si}{\Sigma}

\newcommand{\g}{\gamma}
\newcommand{\vs}[1]{\vspace{#1 mm}}
\newcommand{\hs}[1]{\hspace{#1 mm}}
\newcommand{\lf}{\left(}
\newcommand{\rg}{\right)}

\begin{document}
\hsize=5.6truein
\vsize=9truein
\hoffset=-.3in
\textheight=8.5truein
\voffset=-.8truein
\baselineskip=.60cm
\rightline{hep-th 9902010}

\vs{20}

\centerline{\large\bf A note on a relation between the Killing spinor
and}
\centerline{\large\bf Einstein equations}
\vs{10}
\centerline{Ali Kaya \footnote{partially supported by Scientific and
Technical Research Council of Turkey (T\"{u}bitak)}
\footnote{also at Department of Physics, Bo\~{g}azici University,
\.{I}stanbul,Turkey.}}
\vs{5}
\centerline{Center for Theoretical Physics, Texas A\& M University,}
\centerline{College Station, Texas 77843, USA.}
\vs{15}
\begin{abstract}

It is shown that, 
under certain conditions, the existence of a Killing spinor on a
bosonic background of a supergravity theory implies that the
Einstein equations are also satisfied. As an application of the theorem,
we obtain a new black fivebrane solution of D=11 supergravity, which
has $K3\times R$ topology and preserves 1/4'th supersymmetries of the
theory.

\end{abstract}
\vs{50}
\pagebreak

In string/M theories, solitons are solutions of the corresponding
supergravity equations of motion (for a review see \cite{df},\cite{jp}).
Generically, they can be interpreted
as $p$-dimensional extended black holes (p-branes) and
characterized by  few constants, like the mass and the charges 
of the antisymmetric tensor fields. The extreme members of
black p-branes, which are obtained when the mass and charge saturate a
Bogomolnyi type of bound,  are supersymmetric and have stable multi-source
generalizations.  A background is supersymmetric, if it is invariant under
some supersymmetry transformations of the theory, whose parameters are
called  Killing spinors. The fact that, the states related to 
a supersymmetric soliton  should be in the spectrum at any coupling of
the theory, increases the potential importance of such solutions
\cite{wo}.\\

In many cases, supersymmetric solitons can be obtained by writing suitable  
ansatzs, which solve the Killing spinor and all 
but the Einstein field equations from the beginning \cite{df}.
Then the Einstein equations are solved to determine unknown functions.
This indicates that there may be a connection between the two equation
systems.  In this paper a theorem is proven which shows that, when  the
ansatz  for the fields and geometry is chosen in a suitable way, 
the existence of a Killing spinor implies that the Einstein equations are
also satisfied. We hope that this observation will be
helpful in finding new and interesting supersymmetric solutions. \\

Let us consider a bosonic background ($g_{MN}$,$F_{MNPQ}$) of D=11
supergravity theory \cite{kj} which obeys:  
\be\label{e} R_{MN}=\frac{1}{3}
(F_{M}{}^{PQR} F_{NPQR} -\frac{1}{12} g_{MN}F^{PQRS}F_{PQRS}), 
\ee
\be\label{fe} 
\nabla_{Q}F^{QMNP}=\frac{1}{(24)^{2}}
\epsilon^{MNPA_{1}...A_{8}}F_{A_{1}..A_{4}} F_{A_{5}..A_{8}}.  \ee 
The linearized Rarita-Schwinger equations on this background may be
written
as:
\be
\label{RS} \Gamma^{MNP}D_{N}\psi_{P} = 0,  
\ee 
where the supercovariant derivative $D_{M}$ is given by
\be
D_{M}=\nabla_{M} +
\frac{1}{144}(\Gamma^{PQRS}{}_{M}- 8 \delta^{P}_{M}\Gamma^{QRS}) 
F_{PQRS}, 
\ee 
and $\nabla_{M}$ is the usual covariant derivative acting
on spinors.
Let us further consider a spin 3/2 field $\psi_{M}$, which is
obtained by the action of the supercovariant derivative on an arbitrary
Majorana spinor $\epsilon$:  
\be\label{F} \psi_{M}=D_{M}\epsilon.
\ee 

An interesting question about this spin 3/2 field
is whether it solves the linearized Rarita-Schwinger equations 
(\ref{RS}). The answer is immediately yes, since an affirmative
statement is equivalent to the invariance of D=11 supergravity at the
linearized fermionic level
\footnote{In the linearized theory, the supervariation of (\ref{RS})
should vanish upon imposing other field equations.}.
To verify this let us insert (\ref{F})
into the field equations (\ref{RS}). After a relatively long but 
straightforward calculation
and  $using$ $only$ the equations (\ref{fe}) for the 4-form field,
(\ref{RS})  becomes
\be \label{gt}
(G_{MN} - T_{MN})\Gamma^{N}\epsilon=0,
\ee
where $G_{MN}=R_{MN}-\frac{1}{2} g_{MN}R$ is the Einstein tensor and
\be
T_{MN}=\f{1}{3}(F_{M}{}^{PQR}F_{NPQR}-\f{1}{8}g_{MN}F^{PQRS}F_{PQRS})
\ee
is the energy momentum tensor. 
The presence of the Einstein tensor in equation (\ref{gt}) is due to the
following identity:
\be\label{id}
\Gamma_{M}{}^{NP}\nabla_{[N}\nabla_{P]}\epsilon =
\f{1}{8}\Gamma_{M}{}^{NP}
R_{NPRQ}\Gamma^{RQ}\epsilon = \f{1}{2}G_{MN}\Gamma^{N}\e  .
\ee
Not surprisingly the terms, which contain 
$\nabla_{M}\epsilon$, are canceled. This is because $D_{M}$ is a
certain connection defined on the spinor bundle. The curvature
$\Omega_{MN}$ of this connection is given by
\be
D_{[M}D_{N]}\e = \Omega_{MN} \e ,
\ee
which certainly does not have a term  containing the derivative of $\e$.  
Equations (\ref{gt}) are satisfied  because
$G_{MN}-T_{MN}=0$ are nothing but the Einstein equations (\ref{e}), which
are obeyed by the  background. Therefore we ensure, once more, the
linearized supersymmetry invariance of D=11 supergravity.\\

Now, we try to use the results of the above calculations to prove
a new theorem. We assume this time that the given bosonic background
$does$ $not$ necessarily solve the Einstein equations (\ref{e}), but it
again obeys  the 4-form field equations (\ref{fe}). We further assume
that there exist a Killing spinor $\epsilon_{0}$,
\be\label{killspin}
D_{M}\epsilon_{0}=0;
\ee 
and the Einstein and the  energy momentum tensors have only $diagonal$
entries 
in an orthonormal basis. Differing from the previous case, for the
spin 3/2 field we make the special choice
$\psi_{M}^{0}=D_{M}\epsilon_{0}$ (which is actually zero) and insert it
into the Rarita-Schwinger
equations (\ref{RS}) (which are trivially satisfied). The same 
calculations which have been done to obtain
(\ref{gt}) from (\ref{RS}),  now implies that (\ref{RS}) becomes:
\be \label{gt2}
(G_{MN} - T_{MN})\Gamma^{N}\epsilon_{0}=0.
\ee
Note that to reach (\ref{gt}) from (\ref{RS}), we have $only$ used the
4-form field equations which  are also satisfied by the new
background. Equations (\ref{gt}) are satisfied  for every 
Majorana spinor $\epsilon$, because that background
is chosen to obey Einstein equations. On the other hand
(\ref{gt2}) is correct only for the Killing spinor $\epsilon_{0}$, since
then the Rarita-Schwinger equations are trivially satisfied. When
the indices of (\ref{gt2}) refer to  the orthonormal basis, in
which by assumption only the diagonal entries of  Einstein and
energy momentum tensors are non-zero, it implies:
\be
G_{MN} - T_{MN}=0,
\ee
which are the Einstein equations (\ref{e}) written in a slightly different
form. Therefore we have proven that:\\

(i) if a chosen background solves the 4-form field equations
(\ref{fe}),\\

(ii) if it is known that there exist a Killing spinor on this   
background,\\

(iii) if there exist an orthonormal basis in which the Einstein and
energy momentum tensors have only diagonal entries;\\

then this background, which preserves some fraction of supersymmetries,
also solves the Einstein equations of D=11 supergravity.\\

The main advantage of the theorem is that, to find
a solution of the second order Einstein equations, one can
instead concentrate on the first
order Killing spinor equations, which are of course  easier to solve.  
It is not too difficult to write non-trivial ansatzs which  solve the
4-form field equations from the beginning. Note that, for many of the
known
black p-brane solutions, assumed  form of the geometry and other fields 
have the property (iii) (see for instance \cite{rg},\cite{hs}).
This is because the metric of a static black hole solution depends on a
single radial coordinate and usually has spherical symmetry 
\footnote{The usual multi-source p-brane solutions 
do not obey the condition (iii), since the metrics depend on all
transverse coordinates and the Ricci tensor has non-diagonal entries.}.
Therefore, it is not a big restriction  to fulfill the conditions
of the theorem.\\

Although explicitly proven for D=11 supergravity,  such a theorem can be
generalized to any supergravity theory. To see this, let us remind that
the expression
\be\label{ep}
\Gamma^{MNP}D_{N}D_{P}\epsilon = 0, 
\ee
where 
\be
D_{M}=\nabla_{M}+....
\ee
is the supercovariant derivative of the  supergravity at hand,
$\epsilon$ is an arbitrary spinor and  fields are evaluated on a
bosonic background
which  obeys the equations of motion, is a generic condition for the
linearized supersymmetry invariance. 
In (\ref{ep}),  derivatives of the metric $g_{MN}$ appear only in
the combination of
Einstein tensor, due to the identity (\ref{id}).
Like in  D=11 case, the $\nabla_{M}\e$ terms should cancel.
Therefore, after imposing all but  Einstein equations, 
(\ref{ep}) should become (\ref{gt}) (with the appropriate energy
momentum tensor of the theory) which is also satisfied upon
imposing Einstein equations; ensuring the linearized supersymmetry
invariance.\\ 

With the information above, now we can generalize the theorem.
Dropping the condition that the
background satisfies Einstein equations but assuming the
existence of a Killing spinor, one obtains, following (\ref{ep}), 
again (\ref{gt}) this time
evaluated for the Killing spinor. With condition  (iii), this implies
the Einstein equations. Therefore, if we replace (i) with\\

($i^{'}$) if the background satisfies all but Einstein equations,\\

then together with conditions (ii) and (iii) above , the conclusions of
the theorem should be valid for all supergravities.\\

Let us now illustrate how the theorem may be used for applications.
Consider
type IIB supergravity in 10-dimensions \cite{sch}. We focus on the metric
$g_{MN}$ and anti-self dual 5-form field $F_{MNPQR}$, and set all other
fields to zero. The field equations which govern the dynamics 
may be written as:

\be\label{einst}
R_{MN}=\f{1}{96}F^{PQRS}{}_{M}F_{PQRSN},
\ee
\be\label{ff}
dF=0, {}*F=-F,
\ee
where * is the usual Hodge dual acting on differential forms. The Killing
spinors, on a bosonic background, satisfy:
\be\label{kl}
\nabla_{M}\e +\f{i}{4\times 480} \Gamma^{NP..Q}\Gamma_{M}F_{NP..Q}\e =0,
\ee
where $\e$ is a chiral spinor $\Gamma_{11} \e =\e$;
$\Gamma_{11}=\Gamma_{\hat{0}}....\Gamma_{\hat{9}}$, 
$\Gamma_{11}^{\dagger}=\Gamma_{11}$, and $\Gamma_{11}^{2}=I$.
For illustrational purposes, it is sufficient to write a brane-like ansatz
for the metric:
\be\label{met}
ds^{2}=A(r)^{2}(-dt^{2}+dx_{1}^{2}+..+dx_{3}^{2})+B(r)^{2}dr^{2}+C(r)^{2}d
\Omega_{5}^{2}(m),
\ee
where $d\Omega_{5}^{2}(m)$ is the usual metric of $S_{5}(m)$;
the 5-sphere with
inverse radius $m$. Note that the functions $A$,$B$ and $C$ depend only on
the radial coordinate $r$. The Ricci tensor has only $diagonal$ entries
in the orthonormal basis $E^{\hat{i}}=Adx^{i}$, $E^{\hat{r}}=Bdr$ and
$E^{\hat{a}}=Ce^{\hat{a}}$, where $x^{i}=t,x_{1},..,x_{3}$;{}
$e^{\hat{a}}$ is an orthonormal basis on $S_{5}(m)$.
When we write the Killing spinor equations, all the spinors 
and hatted quantities will refer to the above  orthonormal
basis. We will also use the
notation that, the indices $\mu ,\nu ,\sigma ,\lambda$ refer to
the coordinates $x^{i}$, $r$ and the indices $a ,b,c,d,e$ refer
to the coordinates of $S_{5}(m)$. \\

The 5-form field may initially be chosen to be a 
constant times the volume form on
$S_{5}(m)$. The anti-self duality requires that, there should also be
non-zero components along $x^{i}$ and $r$ directions. As a result one may
write:
\be\label{19}
F_{\mu\nu\sigma\lambda\beta}=\f{-q}{C^{5}}\e_{\mu\nu\sigma\lambda\beta},
\ee
\be\label{20}
F_{abcde}=\f{q}{C^{5}}\e_{abcde},
\ee
where $q$ is a constant and the presence of function $C$ is due to the
fact that the volume forms in (\ref{19}) and (\ref{20})
are determined by the
$metric$ (\ref{met}). The field equations (\ref{ff}) for
$F_{MNPQR}$ are trivially satisfied. The background has now the properties 
$(i^{'})$ and $(iii)$ above.
The parameter $q$ is related to the electric and magnetic charges of the
background, which are identical due to the anti-self duality condition.
On the other hand the Einstein equations (\ref{einst}) become:
\be\label{e1}
-3\left(
\f{A^{'}}{AB}\right)^{2}
-\f{1}{AB}\lf\f{A^{'}}{B}\rg^{'}  -5\f{A^{'}C^{'}}{ACB^{2}}
=-\f{q^{2}}{4C^{10}},
\ee
\be
-\f{4}{AB}\lf\f{A^{'}}{B}\rg^{'} -\f{5}{BC}\lf\f{C^{'}}{B}\rg^{'} = 
-\f{q^{2}}{4C^{10}},
\ee
\be\label{e4}
\f{4a^{2}}{C^{2}}-4\lf\f{C^{'}}{BC}\rg^{2}
-4\f{A^{'}C^{'}}{ACB^{2}}
-\f{1}{BC}\lf\f{C^{'}}{B}\rg^{'}= \f{q^{2}}{4C^{10}},
\ee
where $'$ denotes the derivative with respect to $r$.\\

Let us now turn to the Killing spinor equations (\ref{kl}). We
make (5+5) splitting of the 10-dimensional gamma-matrices $\Gamma_{M}$:
\be
\Gamma_{\mu} = (\g_{\mu}\times I \times \sigma_{1}),
\ee
\be
\Gamma_{a} = (I\times \Si_{a}\times\sigma_{2}),
\ee
where $\g_{\mu}$ and $\Si_{a}$ are 5-dimensional gamma matrices and
$\sigma_{1},\sigma_{2}$ are the usual Pauli matrices. 
The chirality condition of the spinors becomes  $\sigma_{3}\e =\e$.
The Killing spinor equations may now be written explicitly as:   
\be\label{3}
\partial_{i}\e\hs{1}
+\hs{1}\f{A^{'}}{2B}\hs{1}\g_{\hat{i}}\hs{1}\g_{\hat{r}}\hs{1}\e
\hs{1}+ \hs{1}\f{qA}{8C^{5}}\hs{1}\g_{\hat{i}}\hs{1}\e =0,
\ee
\be
\partial_{r}\e\hs{1} +\hs{1} \f{qB}{8C^{5}}\hs{1}\g_{\hat{r}}\hs{1}\e =0,
\ee
\be\label{5}
\nabla_{a}\e \hs{1}- \hs{1}\f{iC^{'}}{2B}
\hs{1}e^{\hat{b}}{}_{a}\hs{1}\Si_{\hat{b}}\hs{1}\g_{\hat{r}}\hs{1}\e 
\hs{1}+ \hs{1}\f{iq}{8C^{4}}\hs{1}e^{\hat{b}}{}_{a}
\hs{1}\Si_{\hat{b}}\hs{1}\e =0,
\ee
where $\nabla_{a}$ is the usual covariant derivative and
$e^{\hat{b}}{}_{a}$ is an orthonormal basis on $S_{5}(m)$. One can choose
$\e$ to be the Killing spinor on $S_{5}(m)$ satisfying:
\be\label{kls}
\nabla_{a}\e + \f{im}{2}e^{\hat{b}}{}_{a}\Si_{\hat{b}}\e = 0.
\ee
Note that the indices on (\ref{5})  and  (\ref{kls}) refer to
different spheres which are  conformally related.
Imposing 
\be 
\g_{\hat{r}}\hs{1}\e\hs{1} =\hs{1} - \e ,
\ee
and
\be\label{1}
\f{A^{'}}{B}=\f{qA}{4C^{5}},
\ee
\be\label{2}
\f{C^{'}}{B}=-\f{q}{4C^{4}} + m,
\ee
the Killing spinor equations become integrable.
Since the background has now  Killing spinors and thus satisfies the
conditions $(i^{'})$,$(ii)$ and $(iii)$, the Einstein equations should
also be trivially satisfied. Indeed, 
one can  check that, when $A,B$ and $C$ obey  (\ref{1}) and
(\ref{2}), they also solve complicated Einstein equations
(\ref{e1})-(\ref{e4}).\\

There are three independent functions and two differential equations which
may be thought to\hs{1} imply that the functions are not constrained
enough.
However, the reparametrization invariance  can be used to set $C=r$. 
Then, from  (\ref{1}) and (\ref{2}), $A$ and $B$ can be solved as:
\bea
B & = & \f{4}{4m - q r^{-4}},\\
A & = & B^{-1/4}.
\eea
This is the  self-dual black threebrane  solution of Duff and Lu
\cite{lu}, written in a slightly different coordinates. To see this, one 
should make the coordinate change $4mr^{4}-q\to r^{4}$ and introduce
Euclidean coordinates for the transverse directions to the threebrane.\\

Note that, this way of solving field equations using Killing spinors
is  different from the methods which were  used in the past in
the context of Kaluza-Klein compactifications, see for instance
\cite{dps}. It is very difficult to apply those methods to
obtain a brane like solution. However, as we have illustrated, it is
very natural to consider Killing spinor equations even to obtain brane
solutions. Here, we have presented an example from IIB supergravity in
10-dimensions, but one can also obtain the M2 and M5 brane
solutions of D=11 supergravity.\\

We now use the theorem to find a new
supersymmetric solution of D=11 supergravity. Basically, one may consider
manifolds having Killing spinors (like anti-de Sitter (AdS) spaces or
spheres) and write suitable  ansatzs involving them. 
One of the simplest possible case  that one may work  is
a space which has covariantly constant spinors like $K3$.
$K3$ is a 4-dimensional Ricci flat K\"{a}hler manifold which has 
self dual (or anti-self dual) curvature tensor. It can be described by
specifying 58 real constants and has no continuous isometries. The metric,
which has self dual curvature, is not explicitly known but its existence
is proven. Unlike a generic, oriented, 4-dimensional manifold, the
holonomy group is just a copy of $SU(2)$ rather than  $SO(4)=SU(2)\times
SU(2)$. This can be seen by the decomposition of $SO(4)$ tensors
with respect to their $SU(2)$ content. The curvature 2-form is an $SO(4)$
valued tensor, however, due to (anti) self duality, it has only non-zero
components in one of the $SU(2)$ bundles. Therefore, the other copy of the 
$SU(2)$ bundle is flat. Because of this property, $K3$ has two linearly
independent covariantly constant $SU(2)$ spinors  which are
left or right handed according to the curvature tensor being self dual or
anti-self dual.\\

Now, consider the following background of D=11 supergravity theory,
\be
ds^{2}= A(r)^{2}(-dt^{2}+dx^{2}+ds^{2}_{K3}) + B(r)^{2} dr^{2} + C(r)^{2}
d\Omega_{4}^{2}(m),
\ee
\be
F\hs{1} = \hs{1}3 \hs{1}q \hs{1}\epsilon_{4},
\ee
where $ds^{2}_{K3}$ is the line element on $K3$, $\epsilon_{4}$ is the
volume form on $S_{4}(m)$, the four sphere with inverse radius $m$,  and
$q$ is a real constant, which is proportional to the the magnetic
charge of the background calculated as the integral of $F$ over 
$S_{4}(m)$.  Note that, as before, the unknown functions $A,B$
and $C$ depend only on the radial coordinate $r$. We choose the following
orthonormal basis on the tangent space, $E^{\hat{i}}=Adx^{i}$,
$E^{\hat{\alpha}}=A e^{\hat{\alpha}}$, $E^{\hat{r}}=Bdr$ and
$E^{\hat{a}}=Ce^{\hat{a}}$, where $x^{i}=(t,x)$, $e^{\hat{\alpha}}$ is an
orthonormal basis on $K3$ and $e^{\hat{a}}$ is an orthonormal basis on
$S_{4}(m)$.  All the hatted indices and spinors will refer to this basis.
We will also use the indices $\mu ,\nu ,\sigma  ...$ to denote  the
coordinates $t,x,r$ and the coordinates of $K3$. The indices $\alpha
,\beta ...$ will refer to the coordinates of $K3$ only.\\

It can easily be checked that the conditions (i) and (iii) of the theorem
are satisfied by the background. Let us therefore consider the Killing
spinor equations. We make the (7+4) splitting of 11-dimensional gamma
matrices $\Gamma_{A}$ as,
\be
\Gamma_{\mu}=(\gamma_{\mu}\times \Sigma _{5}),
\ee
\be
\Gamma_{a}= (I\times \Sigma_{a}),
\ee
where $\gamma_{\mu}$ and $\Sigma_{a}$ are the 7 and 4-dimensional gamma
matrices, respectively. The Killing spinor equations (\ref{killspin}) can
now be written explicitly as,
\be\label{kill1}
\partial_{i}\e\hs{1}
+\hs{1}\f{A^{'}}{2B}\hs{1}\g_{\hat{i}}\hs{1}\g_{\hat{r}}\hs{1}\e
\hs{1}+ \hs{1}\f{qA}{2C^{4}}\hs{1}\g_{\hat{i}}\hs{1}\e =0,
\ee
\be
\nabla_{\alpha}\e\hs{1}
+\hs{1}\f{A^{'}}{2B}\hs{1}
e^{\hat{\beta}}{}_{\alpha}\hs{1}\g_{\hat{\beta}}\hs{1}\g_{\hat{r}}\hs{1}\e
\hs{1}+ \hs{1}\f{qA}{2C^{4}}\hs{1} 
e^{\hat{\beta}}{}_{\alpha}\hs{1}\g_{\hat{\beta}}\hs{1}\e =0,
\ee
\be
\partial_{r}\e\hs{1} +\hs{1} \f{qB}{2C^{4}}\hs{1}\g_{\hat{r}}\hs{1}\e =0,
\ee
\be\label{kill2}
\nabla_{a}\e \hs{1} + \hs{1}\f{C^{'}}{2B}
\hs{1}e^{\hat{b}}{}_{a}\hs{1}\Si_{\hat{b}}\hs{1}\Sigma_{5}\hs{1} 
\g_{\hat{r}}\hs{1}\e  \hs{1}- \hs{1}\f{q}{C^{3}}\hs{1}e^{\hat{b}}{}_{a}
\hs{1}\Si_{\hat{b}}\hs{1}\Sigma_{5}\hs{1}\e =0,
\ee
where $\nabla_{\alpha}$ and $\nabla_{a}$ are covariant derivatives on $K3$
and $S_{4}(m)$, respectively, and $'$ denotes differentiation with respect
to $r$. One can choose $\e$ to be the tensor product of covariantly 
constant spinors on $K3$ and the Killing spinors on $S_{4}(m)$.
Furthermore, it can also be chosen to be independent of the coordinates
$t$ and $x$, and thus satisfies,
\be\label{l1}
\partial_{i}\hs{1}\e \hs{1}= 0,
\ee
\be\label{l2}
\nabla_{\alpha} \e =0,
\ee
\be\label{k2}
\nabla_{a}\e -\f{m}{2} e^{\hat{b}}{}_{a}\hs{1}\Sigma_{\hat{b}}
\hs{1}\Sigma_{5} \hs{1} \e \hs{1} = 0.
\ee
Note the difference between the Killing spinor equations (\ref{kls}) on
$S_{5}$  and (\ref{k2}) on $S_{4}$. Equations (\ref{kill1})-(\ref{kill2})
can  be integrated if one also imposes,
\be\label{proj}
\g_{\hat{r}}\hs{1}\e \hs{1}=\hs{1} -\e,
\ee
and 
\be
\frac{A'}{B}=\frac{q A}{C^{4}},
\ee
\be
\frac{C'}{B}= m - \frac{2q}{C^{3}}.
\ee   
The projection imposed on $\e$ is consistent with the equations
(\ref{l1}),(\ref{l2}) and (\ref{k2}).
Using the reparametrization invariance, one can set $C=r$. 
Solving for $A$ and $B$ we obtain the following 
supersymmetric solution,
\be
ds^{2}=\left(m -\frac{2q}{r^{3}}\right)^{1/3}(-dt^{2}+dx^{2}+ds^{2}_{K3})
+ \left(m -\frac{2q}{r^{3}}\right)^{-2}dr^{2} + r^{2} d\Omega_{4}^{2}(m),
\ee
\be
F\hs{1}=\hs{1}3\hs{1}q\hs{1}\e_{4},
\ee
which represents a magnetically charged fivebrane wrapped on the
manifold $K3$.  The horizon is located at $r=(2q/m)^{1/3}$ and, unlike the
horizon of the well known G\"{u}ven's fivebrane \cite{rg}, it seems to be
a singular surface. One faces with the same kind of singularities  
when the fivebrane wraps a torus \cite{gibb}. At spatial infinity, i.e.
when $r\to \infty$, the solution approaches $R^{7}\times K3$ and therefore
related to the M-theory compactified on $K3$. The mass can be calculated 
with respect to this background and turns out to be proportional to $q$. 
Due to supersymmetry, the mass and the charge saturate the Bogomolnyi 
bound and thus they are equal.\\

There are 8 Killing spinors, which can be constructed as the tensor
product of the single Killing spinor of the space parametrized by
$t,x,r$ obeying (\ref{proj}), 2 covariantly constant spinors of $K3$  
and 4  Killing spinors of $S_{4}(m)$. Therefore, the solution preserves
1/4'th supersymmetries of the theory. \\

As a consequence of the supersymmetry, the new solution we obtained has
also multi-source generalizations \hs{1} which can be found
by defining a new
radial parameter $R$, $R^{3}=mr^{3}-2q$, and introducing cartesian
coordinates, $\vec{R}$, on the Euclidean space transverse to the
fivebranes. In this coordinate system, the fields of multi-source 
fivebranes having $K3\times R$ topology can be written as
\be 
ds^{2}= H^{-1/3}(-dt^{2}+dx^{2}+ds^{2}_{K3})+ H^{2/3} d\vec{R}.d\vec{R},
\ee
\be
F=*dH,
\ee
where $*$ is the hodge dual in the transverse Euclidean space,
$H$ is a generic harmonic function  of the coordinates $\vec{R}$,
\be
H=1+\Sigma_{i} \frac{q_{i}}{|\vec{R}-\vec{R_{i}}|^{3}}
\ee
and the constants $\vec{R_{i}}$ and $q_{i}$ denote  the position and the
magnetic charge  of the $i$'th fivebrane, respectively.\\

This solution is very similar to the G\"{u}ven's multi-source fivebrane,
the only difference is that the 4 flat directions of the branes are
replaced
by $K3$. This amounts to break 16 supersymmetries of the original solution
by another half. Indeed, $K3$ can be replaced by any other Ricci flat four
dimensional manifold, however, a generic choice may break all the
supersymmetries. The presence of $K3$ also breaks the translational
invariance along 4 directions on the branes (note that $K3$ has no
isometries). The effective theory describing the dynamics is a 
$d=2$, $N=8$ supersymmetric field theory. There are 24 fermionic and
12 bosonic zero modes which give 12 on-shell fermionic and bosonic degrees
of freedom. The 9 of the bosonic zero modes are related to the broken
translational invariance and 3 of them should come from the fluctuations
of the four-form field on the fivebrane background.\\

Let us conclude by mentioning  some other possible new  solutions which
may be obtained by using the Killing spinor equations. A
simple modification is to replace the 5-sphere $S_{5}$ in
(\ref{met}) by $S_{2}\times S_{3}$, $S_{1}\times S_{4}$ or
by any other compact space which have Killing spinors and diagonal Ricci
tensors. This may lead to threebrane solutions with different transverse
spaces. Of course, this modification can be applied to all brane
solutions. In some supergravities, one can also work with
Calabi-Yau or other K\"{a}hler manifolds  instead of spheres, which may
give generalizations of compactifications studied in \cite{ca}.\\

Another interesting choice is  conformally AdS or  de-Sitter type
of geometries for the branes.  When the conformal factors depend on a
single radial coordinate transverse to the branes,  these spaces will also
have Killing spinors and diagonal Ricci tensors. This may give branes with
AdS or de-Sitter topologies. One can also try to obtain time dependent
cosmological type  solutions, by changing the role played by the time
coordinate $t$ and the radial coordinate $r$ in the above construction.
These new possibilities  will be studied elsewhere.

\subsection*{Acknowledgements}
I would like to thank S. De\~{g}er for reading the manuscript and helpful
comments.

\end{document}